\documentclass[12pt]{article}
\usepackage{graphicx}
\usepackage{cite}
\usepackage{amsmath} 
\usepackage{hyperref}
\input{epsf} 
\def\ltap{\raisebox{-.6ex}{\rlap{$\,\sim\,$}} \raisebox{.4ex}{$\,<\,$}} 
\def\gtap{\raisebox{-.6ex}{\rlap{$\,\sim\,$}} \raisebox{.4ex}{$\,>\,$}}

\newcommand\as{\alpha_{\mathrm{S}}} 
 
\def\beq{\begin{equation}} 
\def\eeq{\end{equation}} 
\def\beeq{\begin{eqnarray}} 
\def\eeeq{\end{eqnarray}} 
 
\def\Hbb{$H\to b{\bar b}$}
\def\to{\rightarrow}

\def\WH{{\it WH}}
\def\ZH{{\it ZH}}
\def\VH{{\it VH}}

\begin{document} 

\begin{titlepage}
\begin{flushright}
IFUM-1031-FT\\
ZU-TH 23/14
\end{flushright}
\renewcommand{\thefootnote}{\fnsymbol{footnote}}
\par \vspace{10mm}

\begin{center}
{\Large \bf Associated \ZH\ production at hadron colliders:}
\\[0.5cm]
{\Large \bf the fully differential NNLO QCD calculation}
\end{center}
\par \vspace{2mm}
\begin{center}
{\bf Giancarlo Ferrera}$^{(a)}$, {\bf Massimiliano Grazzini}$^{(b)}$\footnote{On leave of absence
from INFN, Sezione di Firenze, Sesto Fiorentino, Florence, Italy.}~~and~~{\bf Francesco Tramontano}$^{(c)}$\\

\vspace{5mm}

$^{(a)}$ Dipartimento di Fisica, Universit\`a di Milano and\\
INFN, Sezione di Milano, I-20133 Milan, Italy\\
$^{(b)}$ Physik Institut, Universit\"at Z\"urich, CH-8057 Z\"urich, Switzerland\\
$^{(c)}$ Dipartimento di Fisica, Universit\`a di Napoli and\\
INFN, Sezione di Napoli, I-80126 Naples, Italy

\vspace{5mm}

\end{center}

\par \vspace{2mm}
\begin{center} {\large \bf Abstract} \end{center}
\begin{quote}
\pretolerance 10000

We consider Standard Model Higgs boson production
in association with a $Z$ boson in hadron collisions. We present
a fully exclusive computation of QCD radiative corrections 
up to next-to-next-to-leading order (NNLO).
Our calculation includes the Higgs boson decay
to bottom quarks ($b$) in next-to-leading order QCD and
the leptonic decay of the $Z$ boson with finite-width effects and spin
correlations. The computation is implemented in a parton level Monte Carlo program
that makes possible to consider arbitrary kinematical cuts on the final-state leptons, the $b$ jets 
and the associated QCD radiation, and to compute the corresponding distributions in the form of bin histograms.
We assess the impact of QCD radiative effects in the boosted kinematics at the LHC and show that
the inclusion of the NNLO corrections is crucial to control the $p_T$ spectrum of the Higgs boson candidate.

\end{quote}

\vspace*{\fill}
\begin{flushleft}
July 2014

\end{flushleft}
\end{titlepage}

\setcounter{footnote}{1}
\renewcommand{\thefootnote}{\fnsymbol{footnote}}


The recent discovery of a neutral boson resonance at the Large Hadron Collider~(LHC)~\cite{Aad:2012tfa,Chatrchyan:2012ufa},
represents the first important step towards the experimental validation of the electroweak symmetry 
breaking mechanism of the Standard Model~(SM).
At the current level of accuracy the data indicate that
this particle has all the
properties of the long sought Higgs boson ($H$)~\cite{Higgs:1964ia,Englert:1964et}, but deviations from the SM predictions are still possible.
Comparisons of theoretical predictions and experimental data in the next years,
can either confirm that the new resonance is indeed
the Higgs boson predicted by the SM or indicate the need for new physics effects.
To this aim it is essential to measure the processes which give information on Higgs boson
couplings to fermions, gauge bosons and its self-coupling and to compare the results 
against the most accurate SM theoretical predictions.

The main production mechanism of the SM Higgs boson at the LHC is the gluon fusion process 
$gg\to H$, through a virtual heavy-quark (mainly top-quark) loop.
This mechanism only provides an indirect evidence for the fermion coupling
to the Higgs boson.
The first direct evidence of such coupling at the LHC is obtained
by the observation of the Higgs boson decay into bottom ($b$) quarks 
or $\tau$ leptons.
Both ATLAS and CMS Collaborations have recently shown evidence for
$H \to \tau \bar{\tau}$ decay~\cite{Chatrchyan:2014nva,ATLAS-CONF-2013-108},
%
while the current experimental situation is not as clear for the $H \to b \bar{b}$
decay.
The problem in the detection of the relatively high rate for $H \to b\bar{b}$
is 
the overwhelming source of background from the QCD direct production
of $b$ jets. 

The associated production of the Higgs boson with a weak gauge boson $V$ ($V=W^\pm,Z$), 
(also known as Higgs-strahlung process) with the vector boson $V$ decaying leptonically, 
provides a clean experimental signature thanks to the presence of a lepton(s) 
with large transverse momentum ($p_T$) and/or large missing transverse energy.
This was the main search channel
for a light Higgs boson at the Tevatron: the combination of CDF and
D0 results lead to the observation of an excess of events,
consistent with the scalar resonance observed at the LHC \cite{Aaltonen:2012qt}.

It has been shown in Ref.~\cite{Butterworth:2008iy} that at the LHC the associated \VH\ production
in the {\it boosted} kinematical regime, 
where the vector boson and/or the Higgs boson have a large $p_T$,
offers the opportunity to disentangle the $H \to b \bar{b}$ signal from backgrounds.
This channel gives also the possibility to separately study the Higgs boson couplings 
to $W$ and $Z$ bosons.

The observation of the associated \VH$(b{\bar b})$ production
at the LHC is however challenging and requires large statistics
at the high centre-of-mass energy of $\sqrt{s}=13/14$~TeV.
At present, with the LHC data at $\sqrt{s}=7/8$~TeV analysed, the CMS experiment observes 
a slight excess of events above the expected SM backgrounds~\cite{Chatrchyan:2013zna} 
while,  on the other hand, the ATLAS experiment observes no significant excess~\cite{Aad:2012gxa,TheATLAScollaboration:2013lia}.

In view of future more precise experimental results that will be available with
the forthcoming LHC run at $\sqrt{s}=13/14$~TeV and with the improvements of the analyses of
the  $\sqrt{s}=7/8$~TeV data samples, it is important to provide accurate theoretical predictions
for cross-sections and differential distributions
in the kinematical region accessed by the experiments.
In Ref.~\cite{Ferrera:2011bk} we presented the computation
of the next-to-next-to-leading order (NNLO) QCD corrections to the fully differential \WH\
hadroproduction and in Ref.~\cite{Ferrera:2013yga} we supplemented such calculation
with the computation of the radiative corrections to the decay
of the Higgs boson into a $b\bar{b}$ pair. 
The corrections to the Higgs
boson decay process 
turn out to be important for the actual experimental analyses of the LHC data at $\sqrt{s}=7/8$~TeV~\cite{Banfi:2012jh}, but well accounted for by the parton shower Monte Carlo.

In this Letter we consider \ZH\ production at hadron colliders and
present, for the first time, a fully differential computation of the NNLO QCD radiative corrections. 
We consider the leptonic decay of the $Z$ boson both to a pair of charged leptons and
to neutrinos and we include finite-width effects and 
spin correlations.
Our calculation is performed by using the $q_T$ subtraction method~\cite{Catani:2007vq},
which applies to the hard-scattering production of colourless high-mass systems in hadron collisions
and it has been successfully  employed  in the computation of NNLO QCD corrections to several 
processes~\cite{Catani:2007vq,Catani:2009sm,Ferrera:2011bk,Catani:2011qz,Grazzini:2013bna,Cascioli:2014yka}.

The status of the higher order QCD prediction for \ZH$(b{\bar b})$ production is the following.
The full NNLO QCD corrections to the total cross-section for \ZH\ production
has been computed in Refs.~\cite{Brein:2003wg,Brein:2011vx} and are available
in the numerical program {\ttfamily vh@nnlo}~\cite{Brein:2012ne}.
Recently also the next-order, i.e.\ ${\cal O}(\as^3)$, 
to the gluon-induced heavy-quark loop mediated subprocess 
has been calculated in Ref.~\cite{Altenkamp:2012sx}
in the limit of infinite top-quark and vanishing bottom-quark masses.
The next-to-leading order (NLO) corrections to \ZH\ production have been implemented at fully differential level in the 
{\ttfamily MCFM} Monte Carlo code~\cite{MCFM}.
An NLO computation matched to the parton shower for $V\!H+1$ jet has been presented in 
Ref.~\cite{Luisoni:2013cuh}, and merged by using the method of 
Ref.~\cite{Hamilton:2012rf}, with the corresponding $V\!H+0$ jet simulation.

Soft-gluon effects to \ZH\ production have been considered in Refs.~\cite{Dawson:2012gs,Li:2014ria}.
The computation of the
fully differential $H\to b{\bar b}$ decay rate in NNLO QCD
has been reported in Ref.~\cite{Anastasiou:2011qx},
while the inclusive $H\to b{\bar b}$ decay rate is
known up to ${\cal O}(\as^4)$~\cite{Baikov:2005rw}.
The electroweak corrections to \ZH\ production and decay have been computed up to NLO accuracy in Ref.~\cite{Ciccolini:2003jy} 
and are included in the fully exclusive numerical program {\ttfamily HAWK}~\cite{Denner:2011id}.

The full NLO and part of the NNLO QCD corrections to \ZH\ production are the same as those of the Drell--Yan (DY) process, in which
the Higgs boson is radiated by the $Z$ boson. In our computation we include the DY-like contributions up to NNLO. 
Besides these contributions, at NNLO
additional non DY-like gluon-induced diagrams have to be considered,
where the Higgs boson couples to a heavy-quark loop \cite{Kniehl:1990iva}.
We have performed an independent computation of these contributions
taking into account the full dependence on the (bottom and top) heavy-quark masses. 
These corrections are substantial at the LHC due to the large gluon 
luminosity, and, as discussed in Ref.~\cite{Englert:2013vua}, they can be particularly relevant in the boosted kinematics.
We have extended the analytical results in Refs.~\cite{Kniehl:1990iva}
to include the decay of the $Z$ and Higgs bosons 
and we checked them numerically with {\ttfamily GoSam}~\cite{Cullen:2011ac}
finding perfect agreement pointwise.
Note that at ${\cal O}(\as^2)$ there is another set of non DY-like contributions
involving quark-induced heavy-quark loops.
These corrections, 
which have been shown to have an impact on the \ZH\ total  cross  section  
at the ${\cal O}(1\%)$  level at the LHC~\cite{Brein:2011vx}, are neglected
in the present paper.
The $H\to b\bar{b}$ decay is computed at NLO by using the dipole subtraction method~\cite{Catani:1996jh} and
it is included at fully differential level both for massless and massive $b$ 
quarks~\footnote{After absorbing the large logarithmic terms of the type $\log(m_H/m_b)$
into the running $Hb\bar{b}$ Yukawa coupling, the effect of the non-vanishing $b$ mass is completely negligible.}. 

By treating the Higgs boson within the narrow width approximation, the differential cross section for 
the associated \ZH$(b{\bar b})$ production and decay processes
can be written as~\footnote{The leptonic decay of the $Z$ boson 
(including spin correlations) has no effect from the point of view of QCD corrections and therefore it has been understood to simplify the notation.}
\begin{equation}
\label{master}
d\sigma_{pp \rightarrow Z\!H+X\to Z b{\bar b}+X} =
\left[
\sum_{k=0}^\infty d\sigma^{(k)}_{pp \rightarrow Z\!H+X}\right] \times
\left[\frac{\sum_{k=0}^\infty d\Gamma^{(k)}_{H \rightarrow b \bar{b}}}
{\sum_{k=0}^\infty \Gamma^{(k)}_{H \rightarrow b \bar{b}}}\right]
\times Br(H \to b \bar{b})\, ,
\end{equation}
where the exponents represent the corresponding order in $\alpha_S$.
In Eq.~(\ref{master}) the fully inclusive QCD effects in the \Hbb\ decay are taken into account by normalising the $Hb{\bar b}$ Yukawa coupling 
such that the value of the Higgs boson branching ratio into $b$ quarks $Br(H\to b{\bar b})$ corresponds to
the precise prediction reported in Ref.~\cite{Dittmaier:2011ti}.

We first consider the NLO corrections both to the production and decay processes. 
Eq.~(\ref{master}) reduces to
\begin{equation}
\label{eqnloproddec}
d\sigma_{pp \rightarrow Z\!H+X\to Z b{\bar b}+X}^{\rm NLO(prod)+NLO(dec)}=
\left[
d\sigma^{(0)}_{pp \rightarrow W\!H} \times
\frac{d\Gamma^{(0)}_{H \rightarrow b \bar{b}}+d\Gamma^{(1)}_{H \rightarrow b \bar{b}}}
{\Gamma^{(0)}_{H \rightarrow b \bar{b}}+\Gamma^{(1)}_{H \rightarrow b \bar{b}}}  +
d\sigma^{(1)}_{pp \rightarrow Z\!H+X}
\times
\frac{d\Gamma^{(0)}_{H \rightarrow b \bar{b}}}
{\Gamma^{(0)}_{H \rightarrow b \bar{b}}}
\right] \times Br(H \rightarrow b \bar{b})\, ,
\end{equation}
which represents the complete NLO calculation because at the first order in $\as$ 
the factorisation between
production and decay is exact due to colour conservation.

When we consider also the NNLO corrections to the production  we have
\begin{align}
\label{eqnnlo}
d\sigma_{pp \rightarrow Z\!H+X\to l\nu b{\bar b}+X}^{\rm NNLO(prod)+NLO(dec)} =&
\left[
d\sigma^{(0)}_{pp \rightarrow Z\!H} \times
\frac{d\Gamma^{(0)}_{H \rightarrow b \bar{b}}+d\Gamma^{(1)}_{H \rightarrow b \bar{b}}}
{\Gamma^{(0)}_{H \rightarrow b \bar{b}}+\Gamma^{(1)}_{H \rightarrow b \bar{b}}} \right. \nonumber \\
&+
\left.
\left(
d\sigma^{(1)}_{pp \rightarrow Z\!H+X}+
d\sigma^{(2)}_{pp \rightarrow Z\!H+X}
\right) \times
\frac{d\Gamma^{(0)}_{H \rightarrow b \bar{b}}}
{\Gamma^{(0)}_{H \rightarrow b \bar{b}}}
\right] \times Br(H \rightarrow b \bar{b})\, .
\end{align}
Although this is not a fully consistent approximation,
since it neglects some ${\cal O}(\as^2)$ contributions in Eq.~(\ref{master}),
it captures the relevant radiative effects.
In particular, thanks to the use of the QCD corrected branching ratio from Ref.~\cite{Dittmaier:2011ti},
if the calculation is sufficiently inclusive over final state QCD radiation, our results fully
include the relevant NNLO effects.
As shown in Ref.~\cite{Ferrera:2013yga}, this is certainly the case for the boosted analysis at $\sqrt{s}=14$ TeV.
Furthermore the calculation allows us, 
for the first time, to assess the impact of the loop-induced $gg$ contribution 
consistently with the other ${\cal O}(\alpha_S^2)$ QCD radiative effects.

In the following we present an illustrative selection of numerical results for \ZH\ production and decay
at the LHC ($pp$ collisions at $\sqrt{s}=8$ and $14$~TeV). 
As for the electroweak couplings, we use the so called $G_\mu$ scheme,
where the input parameters are $G_F$, $m_Z$, $m_W$. In particular we 
use the values
$G_F = 1.16637\times 10^{-5}$~GeV$^{-2}$,
$m_Z = 91.1876$~GeV, $m_W = 80.399$~GeV,
$\Gamma_Z=2.4952$~GeV, $m_t=172$~GeV and $m_b=4.75$~GeV.
The mass of the SM Higgs boson is set to $m_H=125$~GeV, 
the width to $\Gamma_H=4.070$ MeV
and we normalise the $Hb{\bar b}$ Yukawa coupling such that the value of the branching ratio is
$Br(H\to b{\bar b})=0.578$ \cite{Dittmaier:2011ti}.
When no cuts are applied, and the $Z$ and $H$ bosons are produced on shell, our numerical results agree with 
those obtained with the program {\tt vh@nnlo}~\cite{Brein:2012ne}.
We use the NNPDF2.3 parton distribution functions (PDFs) 
sets~\cite{Ball:2012cx}, with
densities and $\as$ evaluated at each corresponding order
(i.e., we use $(n+1)$-loop $\as$ at N$^n$LO, with $n=0,1,2$) and with $\as(m_Z)=0.119$.
The central values of the renormalisation and factorisation scales are fixed to the value 
$\mu_R=\mu_F=m_Z+m_H$ while the renormalisation scale for the 
$H\to b{\bar b}$ coupling is set to the value $\mu_r=m_H$.
The scale uncertainties are computed as follows: we keep $\mu_r=m_H$ fixed and
vary $\mu_R$ and $\mu_F$ independently in the range  
$(m_H+m_Z)/2 \leq \{\mu_R,\mu_F\}\leq 2(m_Z+m_W)$,
with the constraint $1/2\leq \mu_R/\mu_F\leq 2$
(such constraint has the purpose of avoiding large
logarithmic contributions of the form $\ln(\mu_R^2/\mu_F^2)$ in the perturbative expansion).
We then keep $\mu_R=\mu_F=m_Z+m_H$ fixed and  
vary the decay renormalisation scale $\mu_r$ between $m_H/2$ and $2m_H$. The final uncertainty is obtained by
taking the envelope of the two (production and decay) scale uncertainties.

\begin{figure}[t]
\begin{center}
\begin{tabular}{cc}
\includegraphics[width=.50\textwidth]{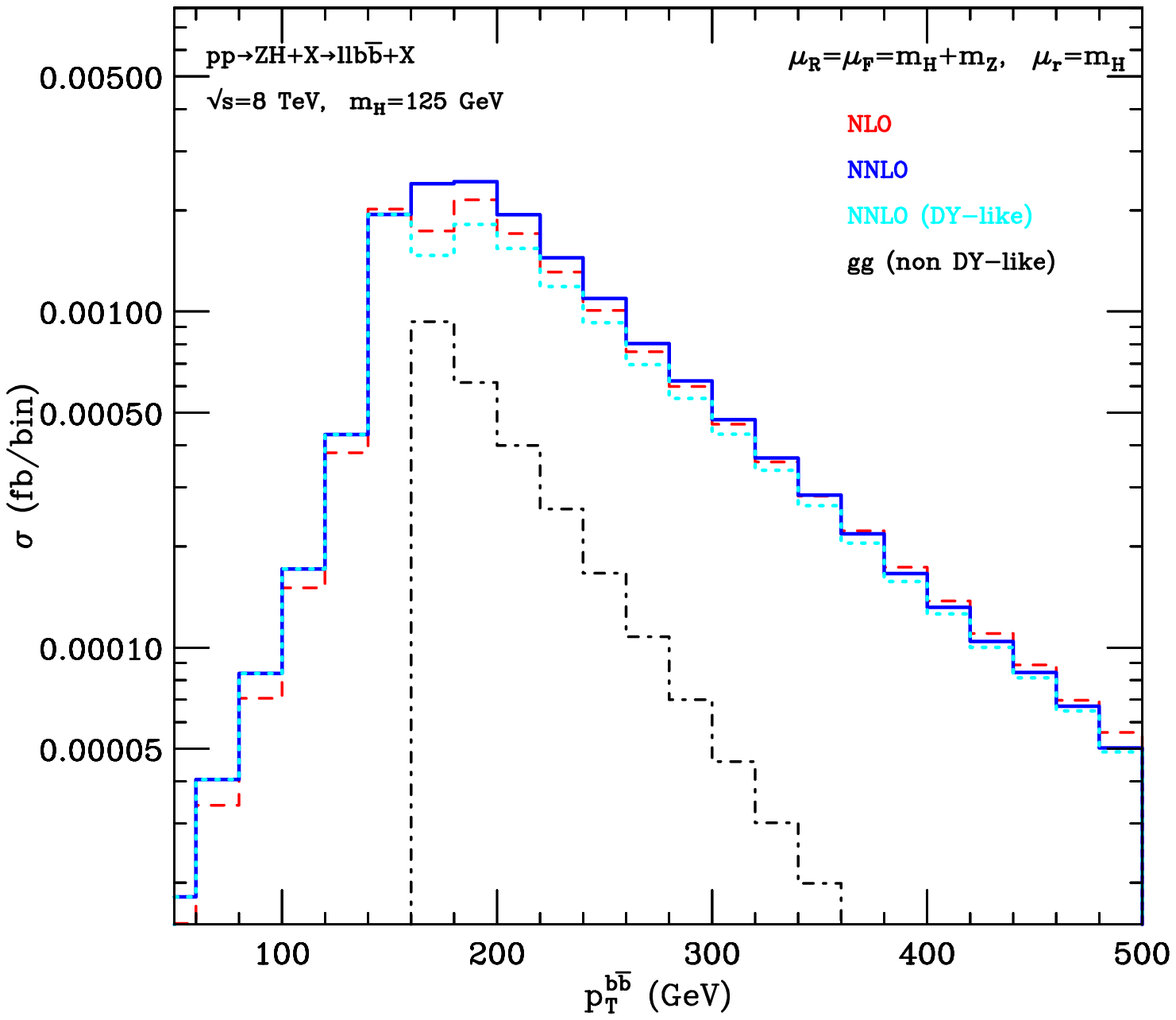}
\includegraphics[width=.452\textwidth]{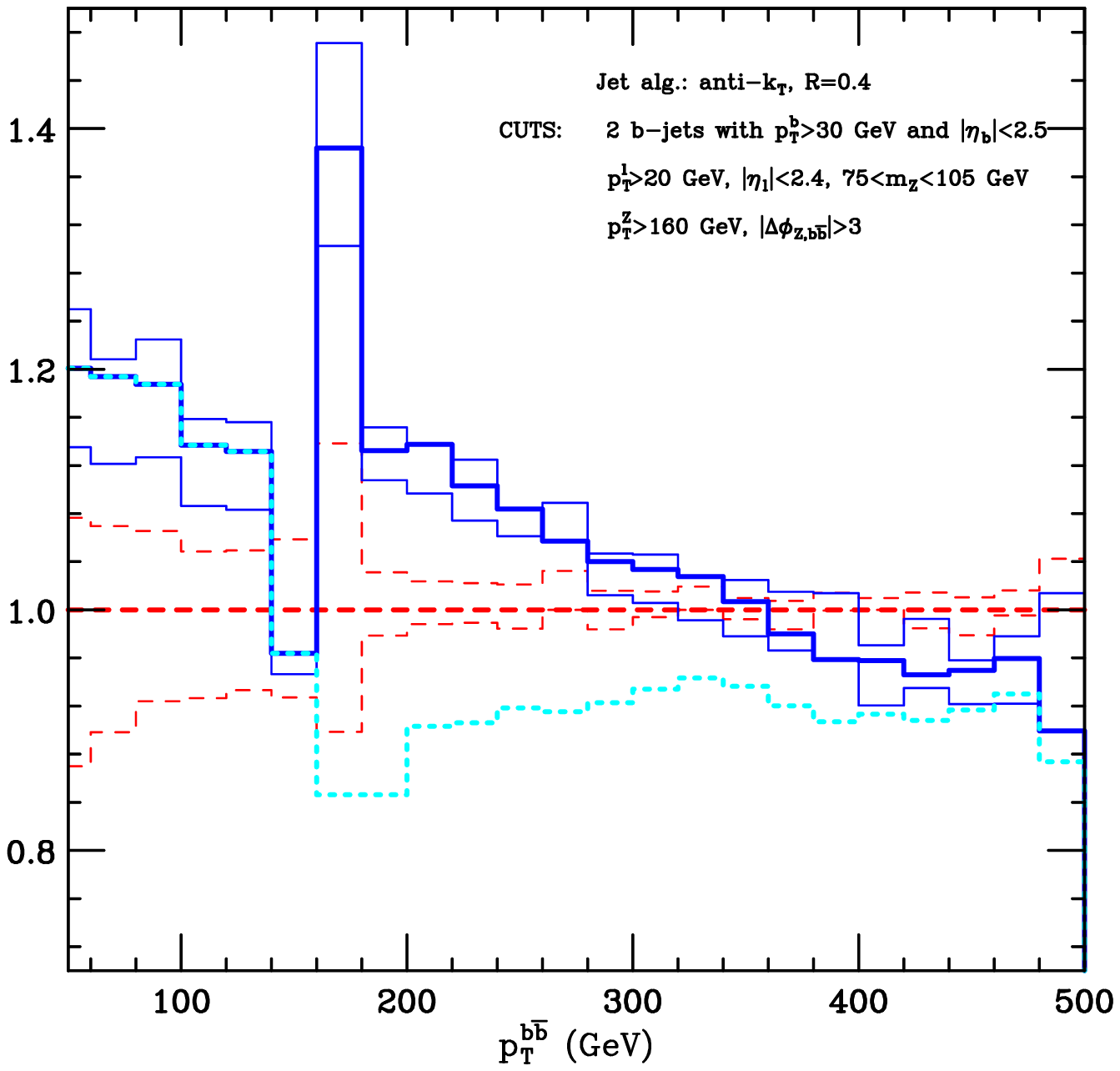}
\end{tabular}
\end{center}
\caption{\label{fig:8cuts}
{\em Left panel: Transverse-momentum distribution of the $b$-jet pair computed at NLO (red dashes), NNLO (blue solid), 
NNLO without the loop-induced $gg$ contribution (cyan dots) and loop-induced $gg$ contribution only (black dot-dashes). 
The applied cuts are described in the text. Right panel: distributions normalised to the NLO result. 
The NLO and NNLO uncertainty bands are also shown. 
}}
\end{figure}

We start the presentation of our results by considering $pp\to ZH+X\to l^+l^-b{\bar b}+X$
at the LHC at $\sqrt{s}=8$~TeV. 
We use the following cuts (see e.g.\ Ref.~\cite{Chatrchyan:2013zna}):
we require the leptons to have transverse momentum $p_T^l > 20$~GeV and pseudorapidity $|\eta_l|< 2.4$, with total invariant mass in the range 
$75-105$~GeV.
The $Z$ boson must have a transverse momentum $p_T^Z > 160$~GeV and
is required to be almost back-to-back with the Higgs boson. To achieve
this condition the azimuthal separation of the $Z$ boson with the $b \bar b$-jet pair must fulfil 
$|\Delta \phi_{Z,b{\bar b}}|>3$. The selection on $p_T^Z$ is important to improve the signal-to-background ratio:
an analogous cut on the Higgs boson can be imposed by focusing
on the region of large transverse-momentum of the $b$-jet pair. 
Jets are reconstructed with the anti-$k_T$ algorithm with $R=0.4$~\cite{Catani:1993hr}:
we require two ($R$) separated $b$-jets each with $p_T^b > 30$~GeV and $|\eta_b|<2.5$.

In Fig.~\ref{fig:8cuts} (left) we study the $p_T^{b\bar b}$ distribution of the  $b$-jet pair from the Higgs boson decay.
We consider QCD predictions at NLO (dashes) and at NNLO (blue), and also show the DY-like NNLO result (dotted) and the loop-induced 
$gg$ contribution (dash-dotted). Both NLO and NNLO results include the NLO corrections to the $H\to b{\bar b}$ decay.
The corresponding
cross sections and scale uncertainties are reported in the first row of Table~\ref{table1}.
In Fig.~\ref{fig:8cuts} (right)
we plot the NLO and NNLO $p_T$ spectra normalised to the full NLO result, together with their scale uncertainty band. 

We see that NNLO DY-like corrections for the production are not negligible:
the accepted cross section is reduced, with respect to NLO, 
by ${\cal O}(10\%)$ with a $K$-factor which is
almost flat in the region $p_T^{b\bar b}\gtap 200$ GeV.
The loop-induced $gg$ contribution has instead a positive 
effect of ${\cal O}(20\%)$ with respect to the NLO result with a $K$-factor
which strongly depends on the $p_T^{b\bar b}$. The overall effect is that 
the NNLO corrections increase the cross section of ${\cal O}(10\%)$.
Given the strong dependence on the transverse momentum, it is crucial that such contribution is properly accounted
for in the experimental analyses.

We observe from Fig.~\ref{fig:8cuts} that NLO and NNLO predictions are affected
by instabilities of Sudakov type~\cite{Catani:1997xc} 
around the LO kinematical boundary $p_T^{b\bar b}\sim 160$~GeV.
As observed in Ref.~\cite{Ferrera:2013yga} the effect of these instabilities can be
reduced by increasing the bin size of the distribution around the critical 
point. 
Moreover the NNLO corrections below the LO kinematical boundary 
($p_T^{b\bar b}\ltap 160$~GeV) are larger, reaching the ${\cal O}(20\%)$ level.
This is not unexpected, since in this region of transverse momenta,
the ${\cal O}(\as)$ result corresponds to a calculation at the first perturbative order and the ${\cal O}(\as^2)$ correction is a next-order term.
The NLO scale uncertainties are ${\cal O}(\pm 10\%)$ in the region 
$p_T^{b\bar b}\ltap 140$ and ${\cal O}(\pm 5\%)$ in the region 
$p_T^{b\bar b}\gtap 200$ GeV and then slightly decrease at higher values of 
$p_T^{b\bar b}$.
The NNLO scale uncertainties is similar in size to the 
NLO one and only partially overlap with the latter.

\begin{figure}[t]
\begin{center}
\begin{tabular}{cc}
\includegraphics[width=.50\textwidth]{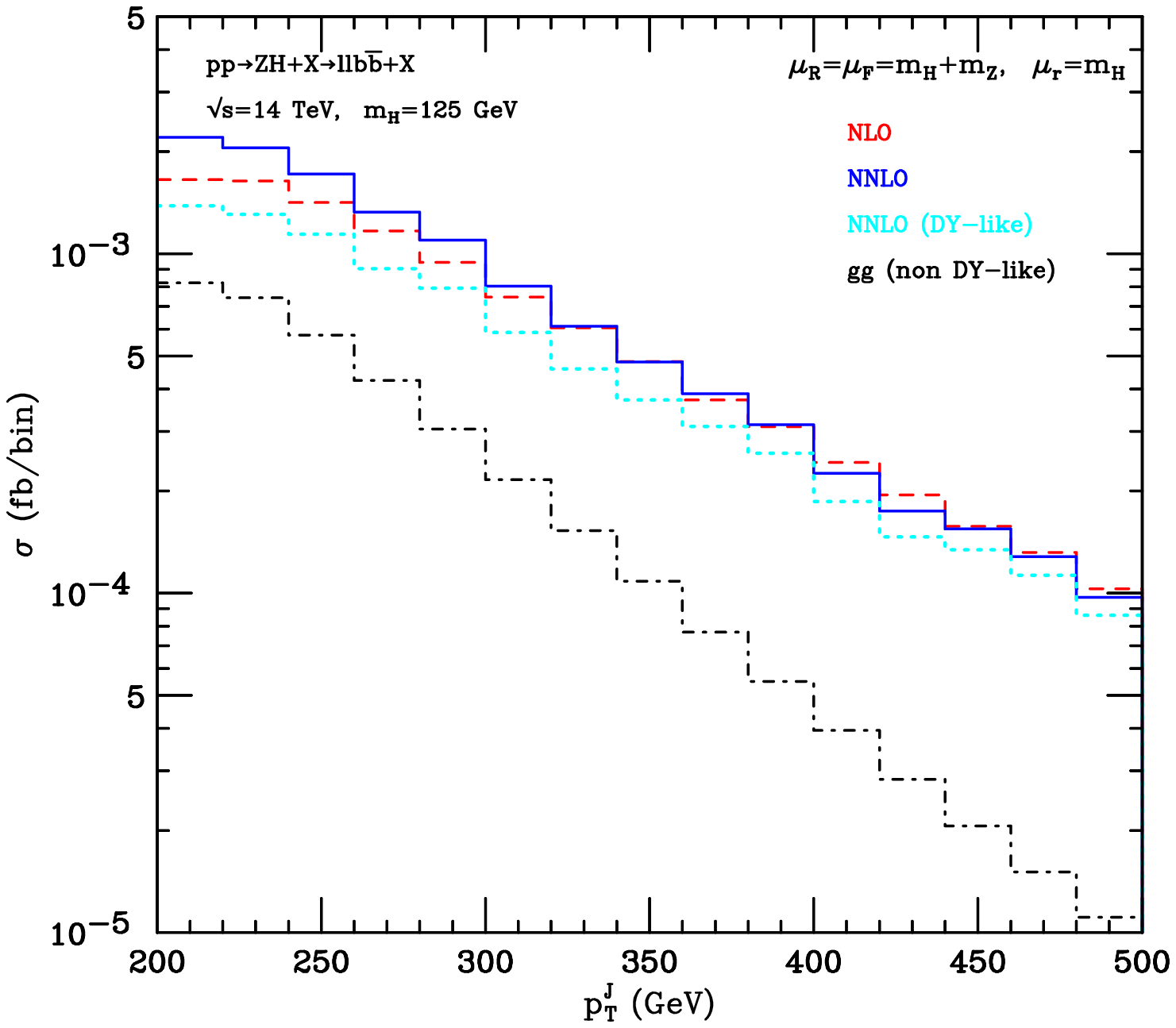}
\includegraphics[width=.46\textwidth]{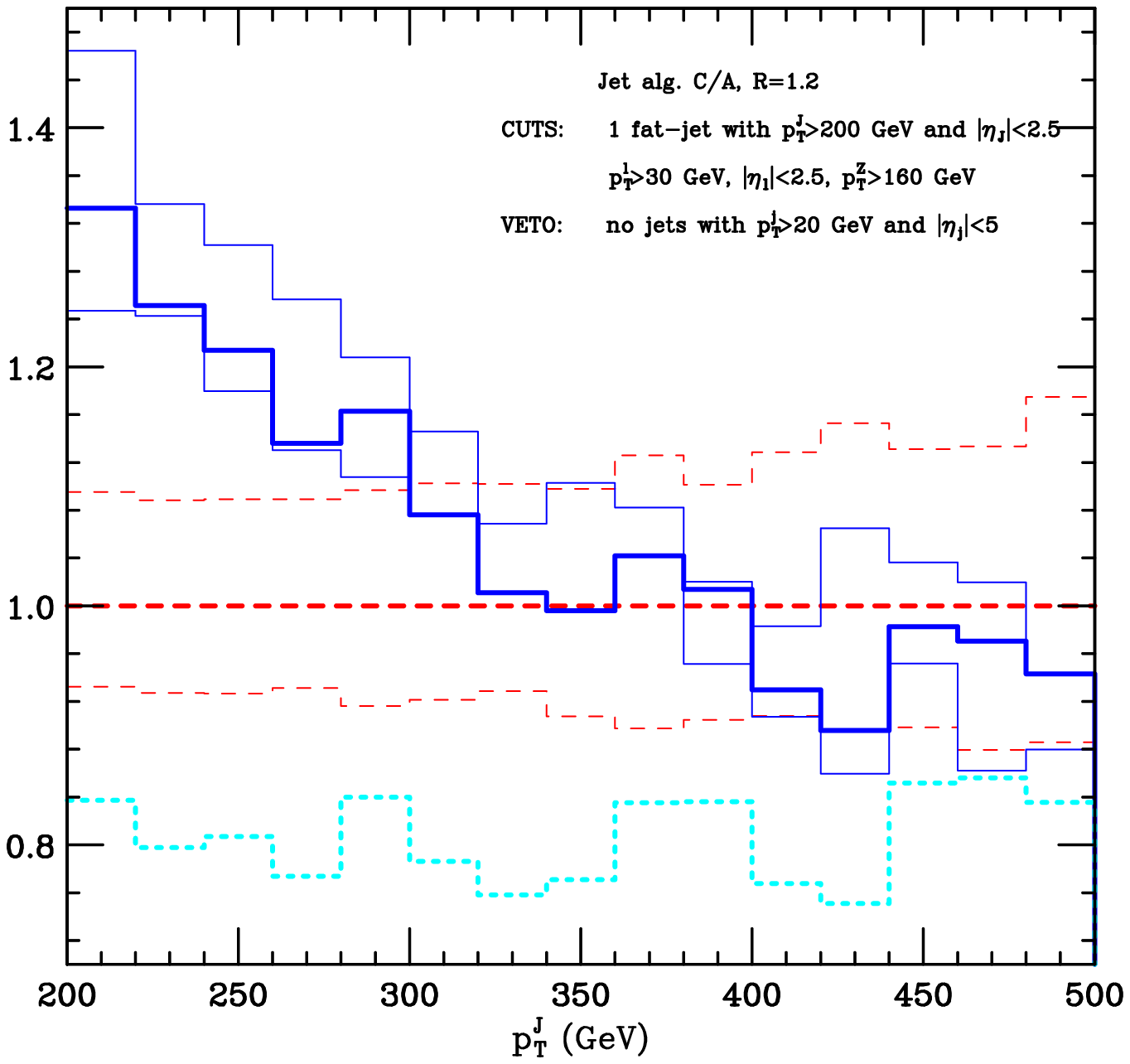}
\end{tabular}
\end{center}
\caption{\label{fig:14fat}
{\em Left panel: Transverse-momentum distribution of the fat jet computed at NLO (red dashes), NNLO (blue solid), NNLO without the loop-induced 
$gg$ contribution (cyan dots) and loop-induced $gg$ contribution only (black dot-dashes). The applied cuts are described in the text. Right panel: distributions normalised to the NLO result. The NLO and NNLO uncertainty bands are also shown.}}
\end{figure}

We next consider $pp\to ZH+X\to l^+l^-b{\bar b}+X$ at the LHC at $\sqrt{s}=14$~TeV. We follow 
the search strategy of Ref.~\cite{Butterworth:2008iy},
where the Higgs boson is
selected at large transverse momenta through its decay into a collimated $b\bar{b}$ pair.
We require the leptons to have $p_T^l > 30$ GeV and $|\eta_l|< 2.5$ with total invariant mass in the range 
$75-105$~GeV.
We also require the $Z$ boson to have $p_T^Z>160$ GeV. 
Jets are reconstructed with the
Cambridge/Aachen algorithm~\cite{Dokshitzer:1997in}, with $R=1.2$.
One of the jets ({\it fat jet}) must have $p_T^J>200$ GeV
and $|\eta_J|<2.5$ and must contain the $b{\bar b}$ pair. 
We also apply a veto on further light jets with $p^j_T>20$ GeV and $|\eta_j|< 5$.
The corresponding
cross sections and scale uncertainties are reported in the second row of Table~\ref{table1}.

The $p_T^J$ distribution of the {\it fat jet} is reported in Fig.~\ref{fig:14fat} (left)
where we consider QCD predictions at NLO and at NNLO for \ZH\ production
with the NLO corrections to the $H\to b{\bar b}$ decay.
In the right panel of Fig.~\ref{fig:14fat}
we plot the $p_T^J$ spectra normalised to the full NLO result together with their scale uncertainty band. 
We have chosen a lower $p_T$ threshold for the $Z$ boson
in order to avoid perturbative instabilities~\cite{Frixione:1997ks,Banfi:2003jj} in
the fixed order predictions around the cut.
The NNLO DY-like corrections for the production are negative and reduce the NLO distribution
by ${\cal O}(20\%)$.
The loop-induced NNLO $gg$ contributions have instead a positive 
impact of about ${\cal O}(25\%)$ which partially compensates the NNLO DY-like corrections to the accepted cross sections.
Nonetheless the behaviour of the overall NNLO corrections strongly 
depends on the
value of $p_T^J$, being positive for $p_T^J\ltap 320$ GeV and
slightly negative for $p_T^J\gtap 400$ GeV.
The NLO and NNLO scale uncertainties bands 
are  ${\cal O}(\pm 10\%)$ and ${\cal O}(\pm 5\%)$ respectively
and they overlap in the region $p_T^J\gtap 280$ GeV .

\begin{table}[htbp]
\begin{center}
\begin{tabular}{|c|c|c|c|}
\hline
$\sigma$ (fb) 
& NLO & NNLO (DY-like) & NNLO \phantom{\big\}} \\
\hline
\hline
LHC8 & $0.2820^{+2\%}_{-2\%}$ 
&  ${0.2574^{+3\%}_{-4\%}}$
&  $0.3112^{+3\%}_{-2\%}$\phantom{\Big\}}\\
\hline
LHC14 & $0.2130^{+10\%}_{-12\%}$
& $ 0.1770^{+7\%}_{-6\%}$
& $ 0.2496^{+5\%}_{-2\%}$\phantom{\Big\}} \\ 
\hline
\end{tabular}
\end{center}
\caption{
{\em Cross sections and their scale uncertainties 
for $pp\to ZH+X\to l^+l^-b{\bar b}+X$ at LHC $\sqrt{s}=8$ and $14$ TeV analyses. The applied cuts are described in the text.
}}
\label{table1}
\end{table}

We add few comments on the uncertainties in our NNLO results.
In Ref.~\cite{Altenkamp:2012sx} the NLO
radiative corrections to the loop-induced $gg$ contribution have been computed by using an effective field theory (EFT) approach.  The validity of the EFT approach to assess the size of these corrections in the boosted regime
is questionable, but the authors of Ref.~\cite{Altenkamp:2012sx} argue that the EFT approach should be reliable if restricted to the computation
of the perturbative correction factor. Under this assumption,
the results of Ref.~\cite{Altenkamp:2012sx} suggest a large impact of radiative corrections, which turn out to be ${\cal O}(100\%)$, thus casting doubts on the convergence of the perturbative expansion. We point out that the NLO corrections to the loop-induced $gg$ contribution are actually only a part of the full N$^3$LO corrections.
Given the large impact of the loop-induced $gg$ diagrams at NNLO, more detailed studies are needed to
precisely assess the theoretical uncertainties in the boosted regime.
At the present stage, we can conclude that the scale uncertainties quoted
in Table 1 cannot be regarded as reliable perturbative uncertainties. A more conservative estimate of the uncertainty can be obtained by comparing the NNLO result to what obtained at the previous order. By taking the difference of the NNLO and NLO results in Table 1 we thus 
obtain an uncertainty of ${\cal O}(\pm 10\%)$ at $\sqrt{s}=8$ TeV and ${\cal O}(\pm 15\%)$ at $\sqrt{s}=14$ TeV.

We have presented the first fully differential
calculation of the cross section for associated \ZH\ production in hadron collisions.
Our calculation accounts for QCD radiative effects to \ZH\ production up to NNLO and includes
QCD effects to the Higgs boson decay up to NLO.
We have studied the impact of radiative corrections in two typical scenarios in $pp$ collisions
at $\sqrt{s}=8$ and $14$~TeV. We have shown
that QCD radiative effects have an important impact on the $p_T$ spectrum of
the Higgs candidate. In particular, the loop-induced $gg$ contribution significantly affects the shape
of the spectrum and should be taken into account in the experimental analyses.
Our calculation is implemented in a parton level Monte Carlo
code that we dub {\ttfamily HVNNLO},
which allows the user to apply arbitrary kinematical cuts on the $Z$ and Higgs decay products as well as on the accompanying QCD
radiation. 
A public version of the {\ttfamily HVNNLO} numerical code, which includes both the associated \ZH\ and \WH\ production processes,  will be 
available in the near future.

\noindent {\bf Acknowledgements.}
We would like to thank Stefano Catani and Andrea Rizzi for helpful discussions.
This research was supported in part by the Swiss National Science Foundation (SNF) under contracts CRSII2-141847, 200021-144352, 
and by the Research Executive Agency (REA) of the European Union under the Grant Agreements PITN--GA---2010-264564 ({\it LHCPhenoNet}), 
PITN--GA--2012--316704 ({\it HiggsTools}). The work of FT is partially supported by MIUR under project 2010YJ2NYW.

\end{document}